\documentclass[a4paper,twocolumn,10pt]{article}

\usepackage{amssymb,amsbsy,psfig,graphicx,times,bbm}
\newcommand{\gr}[1]{\boldsymbol{#1}}
\newcommand{\be}{\begin{equation}}
\newcommand{\ee}{\end{equation}}
\newcommand{\bea}{\begin{eqnarray}}
\newcommand{\eea}{\end{eqnarray}}
\newcommand{\ket}[1]{|#1\rangle}
\newcommand{\bra}[1]{\langle#1|}

\newcommand{\eq}[1]{Eq.~(\ref{#1})}
\newcommand{\sirsection}[1]{\section{\large \sf \textbf{#1}}}

\newcommand{\ack}{\subsection*{\normalsize \sf \textbf{Acknowledgements}}}

\begin{document}

\title{\sf \textbf{\Large Decoherence of number states in phase-sensitive reservoirs}}
\author{Alessio Serafini, Fabrizio Illuminati 
and Silvio De Siena \vspace*{.4cm}\\
\footnotesize Dipartimento di Fisica ``E.~R.~Caianiello'',
Universit\`a di Salerno, INFM UdR Salerno,\vspace*{-.1cm}\\ 
\footnotesize INFN Sez.~Napoli, 
Gruppo Collegato di Salerno,
Via S. Allende, 84081 Baronissi (SA), Italy}

\date{\footnotesize \sf \begin{quote}
\hspace*{.2cm} The non-unitary evolution of initial 
number states in general Gaussian 
environments is solved analytically. Decoherence in
the channels is quantified by determining explicitly the purity 
of the state at any time. 
The influence of the squeezing of the bath on decoherence 
is discussed. The behavior of coherent superpositions of 
number states is addressed as well.
\end{quote}
December 12, 2003}

\maketitle
%%%%%%%%%%%%%%%%%%%%%%%%%%%%%%%%%%%%%%%%%%%%%%%%%%%%%%%
%%%%%%%%%%%%%%%%%%%%%%%%%%%%%%%%%%%%%%%%%%%%%%%%%%%

\sirsection{Introduction}
Recent developments in 
experimental cavity Quantum Electrodynamics
and controlled atom-photon interactions 
allow for a direct investigation of 
deeply quantum mechanical configurations 
of the light field. 
In particular, the {\em deterministic} production of 
low-order number states seem to be 
at hand, by means of micromaser techniques 
in high-$Q$ cavities \cite{walther} 
and of strong coupling with trapped atoms \cite{brown2003}. 
The {\em probabilistic} generation of number states
via conditional measurements and post-selection \cite{kurizki} 
has been demonstrated as well, by 
exploiting parametric down conversion
and low multiplication noise detectors \cite{waks2003}.
A further possibility to generate
number states 
with high fidelities by atom-field interactions in 
high-$Q$ cavities 
has been recently suggested 
\cite{serra}. 
We also mention that, as for motional degrees of freedom, 
effective techniques to create number states 
have been developed and mastered \cite{wineland}.
Moreover, the upcoming VLPC (`visible light 
photon counter') technology holds promising 
perspectives about the actual possibility of 
selectively detect low-order number states \cite{vlpc}.\par
Besides being probes of fundamental
quantum mechanical features, Fock states 
of the electromagnetic field are needed
in several quantum information applications 
whenever, for instance, reliable single--photon 
pulses are required \cite{crypto, secom}.
These possibilities bring to the attention 
the problem of preserving the 
quantum mechanical properties of number states, 
which are unavoidably corrupted by environmental 
decoherence. Indeed, their inherent 
non--classical nature makes such states 
especially fragile and difficult to maintain. 
More specifically, the numerical analysis strongly 
suggests that the very possibility of generating 
pure number states is seriously hurdled by environmental
decoherence, even in high-$Q$ cavity settings \cite{nayak}.
In this paper we study the rate of decoherence 
of initial number states in general Gaussian noisy channels,   
{\em i.e.~}in presence of dissipation in general 
Gaussian environments. 
The dynamical behavior of the system is described   
by the quantum optical master equation, 
allowing for arbitrary phase--sensitive     
(`squeezed') baths.  
The dynamics will be solved in terms of the 
symmetrically ordered characteristic function, while
decoherence will be quantified by 
computing the purity $\mu=\,{\rm Tr}\,\varrho^2$  
during the evolution of the state. 
%%%%%%%%%%%%%%%%%%%%%%%%%%%%%%%%%%%%%%%%%%%%%%%%%5
\sirsection{Solving the master equation}
Let us consider a denumerable, infinite-dimensional
Hilbert space ${\cal H}$, 
spanned by a Fock basis $\{\ket{n}\}$, with $n\in \mathbb{N}$, 
of eigenstates of the hermitian operator $\hat{n}=a^{\dag}a$. 
The annihilation and creation operators $a$ and $a^{\dag}$ 
satisfy the canonical commutation relation $[a,a^{\dag}]={\mathbbm 1}$. 
We define the quadrature operators $\hat{x}=(a+a^{\dag})/\sqrt{2}$ 
and $\hat{p}=-i(a-a^{\dag})/\sqrt{2}$ describing, for instance, 
amplitude and phase quadratures of a single mode of the 
electromagnetic field, or 
position and momentum operators of a material 
harmonic oscillator.\par
Any quantum state of this system can be described 
either by its density matrix $\varrho$ or by
its symmetrically 
ordered characteristic function $\chi(\alpha)$ \cite{barnett}, defined as
\be
\chi(\alpha)=\,{\rm Tr}(\varrho D_{\alpha}) \; ,
\ee
where $D_{\alpha}=\,{\rm exp}(\alpha a^{\dag}-\alpha^{*}a)$ is 
the unitary displacement operator. In the following 
we will make use of phase space variables $x$ and $p$, defined by 
$\alpha=(x+ip)/\sqrt{2}$.
Moreover, it is useful to define the covariance 
matrix $\gr{\sigma}$, associated to a state $\varrho$ by
$$
\sigma_{ij}=\frac{1}{2}\langle \hat{x}_i \hat{x}_j + 
\hat{x}_j \hat{x}_i \rangle -
\langle \hat{x}_i \rangle \langle \hat{x}_j \rangle \, ,
$$ 
with $\hat{x}_1 = \hat{x}, \hat{x}_2 = \hat{p}$ and 
$\langle O\rangle=\,{\rm Tr}\,(\varrho O)$ for the operator $O$.
The dynamics we will study can be modeled by the coupling
with a continuum of oscillators, described 
by the following interaction Hamiltonian
\be
H_{int}=\hbar\int [W(\omega)a^{\dag}b(\omega)+W(\omega)^{*}ab^{\dag}]
\,{\rm d}\,\omega \, ,
\ee
where $b(\omega)$ stands for the annihilation operator of the bath mode 
labeled by the variable $\omega$, whereas $W(\omega)$  
represents the coupling. The state of the bath has been assumed to be 
stationary.
Under the Markovian approximation, such a coupling
gives rise to a time evolution ruled by the following 
master equation (in the interaction picture) \cite{gardiner}
\bea
\dot\varrho & = & \frac{\gamma }{2}\Big(N \: L[a^{\dag}]\varrho
+(N+1)\:L[a]\varrho - \nonumber \\
&&\:  M^{*}\:D[a]\varrho + M
\:D[a^{\dag}]\varrho \Big)
\label{rhoev} \, ,
\eea
where the dot stands
for time--derivative,
the Lindblad superoperators are defined as
$L[O]\varrho \equiv  2 O\varrho O^{\dag} -
O^{\dag} O\varrho -\varrho O^{\dag} O$ and
$D[O]\varrho \equiv  2 O\varrho O -O O\varrho -\varrho O O$,
the coupling is $\gamma=2\pi W^{2}(0)$,
while the coefficients $N$ and $M$ 
are defined in terms of the correlation functions 
$\langle b^{\dag}(0)b(\omega) \rangle = N\delta(\omega)$ and 
$\langle b(0)b(\omega) \rangle = M\delta(\omega)$, 
where averages are computed over the state of the bath.
The requirement of positivity of the density matrix imposes
the constraint $|M|^{2} \le N(N+1)$.
At thermal equilibrium, {\it i.e.}~for $M=0$, $N$
coincides with the average number of thermal photons
in the bath.
If $M\neq 0$ then the bath is said to be `squeezed', or phase-sensitive, 
entailing reduced fluctuations in one field quadrature. 
A squeezed reservoir 
may be modeled as the interaction with a bath of 
oscillators excited in squeezed thermal states \cite{sqbath1}; 
several effective realization of such reservoirs have been 
proposed in recent years \cite{sqbath}.\par 
In general, the real parameter $N$ and the complex
parameter $M$ allow for the description 
of the most general single--mode Gaussian reservoir, 
fully characterized by its covariance 
matrix $\gr{\sigma}_{\infty}$, given by 
\be
\gr{\sigma}_{\infty}=\left(\begin{array}{cc}
\frac12+N+\,{\rm Re}\,M & {\rm Im}\,M \\
{\rm Im}\,M & \frac12+N+\,{\rm Re}\,M
\end{array}\right) \; . \label{envi}
\ee 
Such a Gaussian state constitutes the asymptotic state 
in the channel, irrespective of the initial condition.
The parameters $\gamma$, $N$ and $M$
completely characterize the Gaussian channel. 
A more suitable parametrization of the environmental 
state is provided by the following equations \cite{paris03}
\begin{eqnarray}
\mu_{\infty}&=&\frac{1}{\sqrt{(2N_{i}+1)^{2}-4|M_{i}|^{2}}} 
\: , \label{purasi} \\
&& \nonumber \\
\cosh(2r)&=&\sqrt{1+4\mu_{\infty}^{2}|M_{i}|^{2}}
\: , \label{squizasi} \\
&& \nonumber \\
\tan(2\varphi)&=&-\tan\left({\rm Arg}{M_{i}}\right)
\: . \label{phiasi}
\end{eqnarray}
The quantities $\mu_{\infty}$, $r$ and 
$\varphi$ are, respectively, the purity, 
the squeezing parameter and the squeezing angle of 
the squeezed thermal state of 
the bath. This parametrization will prove useful
in the following.\par
Eq.~(\ref{rhoev}) is equivalent to the following 
diffusion equation for the characteristic function 
$\chi$ in terms of the quadrature variables $x$ and $p$
\be
\dot{\chi}  =  -\frac{\gamma}{2}\Bigg[
(x \; p){\partial_{x} \choose \partial_p}\chi
+(x \; p) \gr{\sigma}_{\infty}{x \choose p}\chi
\Bigg] \, .
\label{diff}
\ee
It is easy to verify that, for any initial condition 
$\bar{\chi}(x,p)$, the following expression solves Eq.~(\ref{diff}) 
\be
\chi=\bar{\chi}(x\,{\rm e}^{-\frac\gamma2 t},p\,{\rm e}^{-\frac\gamma2 t})
\,{\rm e}^{-\frac12(x\; p)\gr{\sigma}_{\infty}{x \choose p}
(1-\,{\rm e}^{-\gamma t})} \; . \label{solu}
\ee
The initial condition we will deal with is a number state 
$\ket{n}\bra{n}$, whose symmetric characteristic 
function $\chi_n$ can be easily determined \cite{barnett}, and reads
\be
\chi_n(\alpha)=\bra{n}D_\alpha\ket{n}=\,{\rm e}^{-\frac{|\alpha|^2}{2}}
L_n (|\alpha|^2) \, , \label{ini}
\ee
where $L_n$ is the Laguerre polynomial of order $n$
\be
L_n (x)=\sum_{m=0}^{n}\frac{(-x)^m}{m!}{n \choose m} \; .
\ee
Putting Eqs.~(\ref{solu}) and (\ref{ini}) together and switching 
again to quadrature variables yields the solution $\chi_n(t)$, 
accounting for the evolution of the initial number state $\ket{n}$
in the noisy channel
\be
\chi_n(t)=L_n\left(\frac{x^2+p^2}{2}\,{\rm e}^{-\gamma t}\right)
\,{\rm e}^{-\frac12 (x\;p)\gr{\sigma}(t){x \choose p}} \, , \label{solun}
\ee
with
\be
\gr{\sigma}(t)=\frac{\mathbbm 1}{2}\,{\rm e}^{-\gamma t}+
\gr{\sigma}_{\infty}(1-\,{\rm e}^{-\gamma t}) \; .
\ee
Notice that, clearly, $\gr{\sigma}(t)$ {\em is not} the covariance 
matrix of the evolving state, because of the presence of 
the Laguerre polynomial in Eq.~(\ref{solun}).
%%%%%%%%%%%%%%%%%%%%%%%%%%%%%%%%%%%%%%%%%%%%%%%%%5
\sirsection{Decoherence of number states}
Decoherence of the initial pure state in the channel
will be quantified by following the evolution of the purity $\mu=\,{\rm Tr}\,\varrho^2$. 
Such a quantity properly describes the degree of mixedness 
of a quantum state $\varrho$. For continuous-variable (CV) systems it
takes the value $1$ for pure states (represented by normalized projectors) 
and the value zero for maximally mixed states. The conjugate 
of $\mu$ is referred to as the `linear entropy' $S_{L}$ in information 
theory: $S_{L}=1-\mu$.\par
The purity of a quantum state of a single-mode CV system
is easily computed as an integral over 
the whole phase space \cite{paris03}
\be
\mu=\frac{1}{2\pi}\int_{\mathbb R}\int_{\mathbb R}|\chi|^2 
\,{\rm d}\,x\,{\rm d}\,p \; . \label{purcar}
\ee
The generalization of Eq.~(\ref{purcar}) to multi-mode systems
is straightforward, and allows to track the dynamics in noisy
channels of entangled two-mode Gaussian states \cite{serafini03}
and of Schr\"{o}dinger cat-like states \cite{cats03}. Moreover, 
for CV systems, it has been recently proved
that knowledge of the global and marginal purities provides
a strong and experimentally reliable characterization of
the entanglement of arbitrary two-mode Gaussian mixed states
\cite{adesso03}.
In the present instance, Eqs.~(\ref{solun}) and (\ref{purcar})  
allow to compute the purity $\mu_n(t)$ of an initially pure number state
of order $n$ evolving in the Gaussian channel.
%%%%%%%%%%%%%%%%%%%%%%%%%%%%%%%%%%%%%%%%%%%%%%%5
\subsection{Thermal bath}
The instance of a reservoir at thermal equilibrium 
corresponds to the choice $M=0$. In this case,
both the environmental Gaussian state and the 
initial condition are 
rotationally invariant in phase space, see Eqs.~(\ref{envi}) and (\ref{ini}), 
so that polar coordinates constitutes a suitable and convenient choice. 
Employing the variable $s=\,{\rm e}^{-\gamma t}|\alpha|^2$, 
one gets
\be
\mu_n(t)=\,{\rm e}^{\gamma t}\int_0^\infty L_n^2(s)
\,{\rm e}^{-\left(\,{\rm e}^{\gamma t}(2N+1)-2N\right)s}\,{\rm d}s \; . 
\ee
Which can be analytically solved, using the relation \cite{gradstein}
\[
\int_0^\infty \,{\rm e}^{-ax}L_n^2 (x) \,{\rm d}\,x=
\frac{(a-2)^n}{a^{n+1}}P_n\left(1+\frac{2}{a^2-2a}\right) \, ,
\]
where $P_n$ is the Legendre polynomial of order $n$ 
\[
P_n(x)=\frac{1}{2^n n!}\frac{{\rm d}^n}{{\rm d}\,x^n}(x^2-1)^n \; ,
\]
to find
\be
\mu_n(t)=\,{\rm e}^{\gamma t}\frac{(\xi-2)^2}{\xi^{n+1}}
P_n\left(1+\frac{2}{\xi^2-2\xi}\right) \, , \label{purter}
\ee
\be
{\rm with}\quad \xi=\,{\rm e}^{\gamma t}(2N+1)-2N \, . \label{csi}
\ee
\eq{purter} provides the exact evolution of the purity 
of an initial number state $\ket{n}$ in a thermal channel, 
fully determined
by its mean photon number $N$.
Quite clearly, the purity in the channel 
is a decreasing function of $N$. 
It is a decreasing function of $n$ as well:  
higher order number states are more fragile 
and decohere faster. Moreover, 
number states with $n > 0$ 
can show a local minimum of 
the purity and a partial revival up to the 
asymptotic purity $\mu_{\infty}=1/(2N+1)$. 
%%%%%%%%%%%%%%%%%%%%%%%%%%%%%%%%%%%%%%%%%%%%%%%%%
\begin{figure}[tb]
\begin{center}
\includegraphics[width=7.5cm]{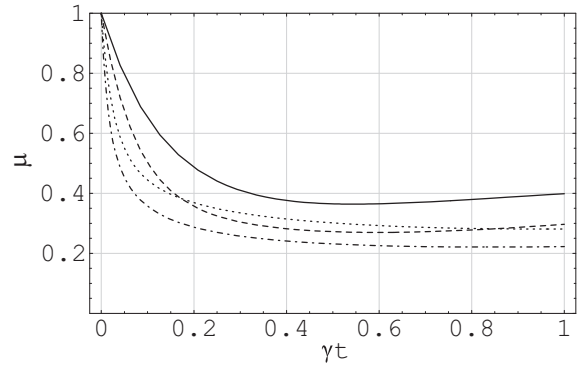}
\caption{\sf\footnotesize Evolution of purity for initial number states in Gaussian channels with
$\mu_{\infty}=0.5$ (purity revivals are evident in the plot). 
The solid line refers to state $\ket{1}$ in a non squeezed bath; 
the dotted line refers to state $\ket{1}$ in a bath with $r=1$; 
the dashed line refers to state $\ket{2}$ in a non squeezed bath; 
the dot--dashed line refers to state $\ket{2}$ in a bath with $r=1$.\label{nums}}
\end{center}
\end{figure}
%%%%%%%%%%%%%%%%%%%%%%%%%%%%%%%%%%%%%%%%%%%%%%%%%5
\subsection{Squeezed bath}
We will now deal with the general instance $M\neq0$.
Due to the rotational symmetry of the  
characteristic functions of number states, we are free to choose 
any direction of squeezing of the environmental state
of the channel. 
Indeed, the rate of decoherence can only depend on the module $|M|$
of the parameter $M$, which can therefore be chosen real and positive
(corresponding to the choice $\varphi=0$),  
without loss of generality.
Making such a choice and exploiting \cite{gradstein}
\[
\int_{0}^{2\pi}\,{\rm e}^{p\cos\varphi}\,{\rm d}\,\varphi=
2\pi I_{0}(|p|)
\]
(where $I_{0}(x)=J_{0}(ix)=
\sum_{k=0}^{\infty}\frac{x^{2k}}{(2^k k!)^2}$ is the  
zero order modified Bessel function of the first kind), one eventually finds 
\be
\mu_{n}(t)=\,{\rm e}^{\gamma t}
\int_{0}^{\infty}\,{\rm e}^{-\xi s}
L_n(s)
I_0\left(|M|(\,{\rm e}^{\gamma t}-1)s\right)\,{\rm d}s \, . \label{pursq}
\ee 
Such an integral cannot be further simplified, 
but can be numerically estimated 
to analyze the effect of the squeezing of the bath on
the decoherence of number states. 
Eq.~(\ref{pursq}) obviously reduces to Eq.~(\ref{purter}) for $M=0$. 
Besides, the asymptotic expressions can be analytically integrated 
to yield 
$$\lim_{t\rightarrow\infty}\mu_n(t)=\mu_{\infty}=
\frac{1}{\sqrt{(sN+1)^2-4|M|^2}}\; ,$$ 
which, according to Eq.~(\ref{purasi}), is the asymptotic purity 
in the channel, fixed by the reservoir state, irrespective of the 
chosen initial condition.\par
\begin{figure}[tb]
\begin{center}
\includegraphics[width=7.5cm]{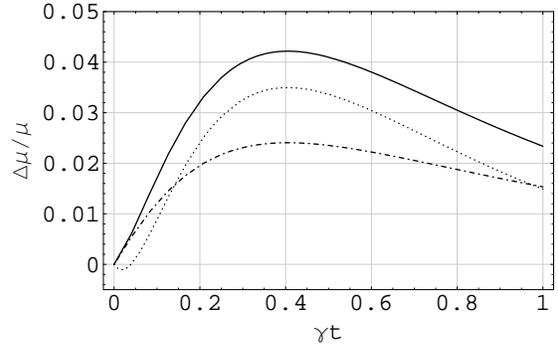}
\caption{\sf \footnotesize The relative increase in purity, defined by $\Delta\mu/\mu=
(\mu_{01}(t,r)-\mu_{01}(t,0))/\mu_{01}(t,0)$, as a function of time during the 
evolution of the superposition $\ket{\psi_{01}}$ in Gaussian channels. 
The optimal condition $\vartheta=\varphi+\pi/2$ is always assumed, while $\mu_{\infty}=0.5$. 
The solid line refers to a bath with $r=0.28$, close to the optimal value; 
the dotted line refers to a bath with $r=0.4$ and the dot--dashed line refers to a bath
with $r=0.1$.\label{cats}}
\end{center}
\end{figure}

Eq.~(\ref{pursq})
shows that $\mu_{n}$ is an increasing function of $|M|$.
However, the dependence on squeezing has to be 
analyzed by considering the parameters $\mu_{\infty}$ and $r$ 
instead of $N$ and $M$, because they permit to study
the effect of squeezing (quantified by $r$) at given
asymptotic purity $\mu_{\infty}$. Such a dependence 
can be reconstructed inserting Eqs.~(\ref{purasi}, \ref{squizasi}, \ref{csi}) 
into Eq.~(\ref{pursq}) and turns out to be quite involved. 
Anyway, a numerical analysis has been carried out, and is
summarized in Fig.~\ref{nums}, where the evolution at short times 
($t\le\gamma$)
is considered. Note that this is the interesting time range, in which decoherence 
takes place before the system is driven towards the 
squeezed thermal state of the environment.
Such an analysis definitely shows that 
squeezing the bath does not slow down the decoherence rate of
number states in Gaussian channels. At a given asymptotic purity 
$\mu_{\infty}$ the highest purity is maintained for $r=0$. 
%%%%%%%%%%%%%%%%%%%%%%%%%%%%%%%%%%%%%%%%%%%%%%%%%5
%\begin{figure}[tb]
%\begin{center}
%\includegraphics[width=7.5cm]{numl.ps}
%\caption{\label{numl}}
%\end{center}
%\end{figure}
%%%%%%%%%%%%%%%%%%%%%%%%%%%%%%%%%%%%%%%%%%%%%%%%%5
\sirsection{Coherent superpositions}
We now consider the behavior of initial 
coherent superpositions of number states evolving in a general 
Gaussian noisy channel. To properly exemplify the 
decoherence of such states, we 
focus on the simplest coherent normalized 
superposition $\ket{\psi_{01}}=(\ket{0}+\,{\rm e}^{i\vartheta}\ket{1})/\sqrt{2}$
(which constitutes a `microscopic Schr\"odinger cat').
The characteristic function $\chi_{01}$ of this state 
is simply found \cite{barnett}
\be
\chi_{01}(\alpha)=\frac{{\rm e}^{-\frac{|\alpha|^2}{2}}}{2}
\left[2-\,{\rm e}^{-\gamma t}|\alpha|^2-\,{\rm e}^{-\frac{\gamma t}{2}}
(\alpha^{*}\,{\rm e}^{-i\vartheta}-\alpha\,{\rm e}^{i\vartheta})\right]  .
\ee
Inserting $\chi_{01}$ as the initial condition in \eq{solu} and 
performing the integration of \eq{purcar} yields, for the 
purity of the initial cat-like state evolving in the channel
\bea
\mu_{01}(t,r)&=&4\nu-\,{\rm e}^{-2\gamma t}\frac{\nu^2}{2\mu_{\infty}}
\Big(\mu_{\infty}+(\,{\rm e}^{\gamma t}-1)(\cosh(2r) \nonumber\\
&+&\cos(2\vartheta-2\varphi)\sinh(2r))\Big)\nonumber\\
&+&\,{\rm e}^{-4\gamma t}\frac{\nu^5}{2\mu_{\infty}^2}
\Big(4\mu_{\infty}^2+8(\,{\rm e}^{\gamma t}-1)\mu_{\infty}\cosh(2r)\nonumber\\
&+&(\,{\rm e}^{\gamma t}-1)^2(3\cosh(4r)+1)\Big)
\label{purcat}\eea
where 
\be
\nu=\bigg[\frac{1}{\mu_{\infty}^{2}}\left(1-
{\rm e}^{-\gamma t}\right)^{2} \, + \,
{\rm e}^{-2\gamma t}
+2\frac{1}{\mu_{\infty}}\cosh(2r)
\bigg]^{-1/2}
\ee
is the purity of an initial vacuum in the channel \cite{paris03}. \eq{purcat}
shows that, quite interestingly, the evolution of the coherent superposition 
is sensitive to the phase $\varphi$ of the bath. It is straightforward 
to see that the optimal choice maximizing purity at any given time 
is provided by $\vartheta=\varphi+\pi/2$. Fixing such a choice, 
we have numerically analyzed the dependence of $\mu_{01}$ on $r$:
for small squeezing parameters, the purity $\mu_{01}$ does increase with $r$. 
The optimal choice for $r$ depends on time, for $\gamma t=0.5$ it turns out to 
be $r\simeq0.28$. The relative increase in purity is plotted 
in Fig.~\ref{cats}, as a function of time, for various choices of the 
squeezing parameter $r$.
%%%%%%%%%%%%%%%%%%%%%%%%%%%%%%%%%%%%%%%%%%%%%%%%%
\sirsection{Comments and Conclusions}
We have analytically solved the non-unitary evolution 
of number states in Gaussian noisy channels, and 
quantitatively estimated their decoherence.
The dissipative model we have considered 
covers a variety of physical situations, from thermal 
dissipation of stationary modes in optical cavities 
to corruption of traveling waves in lossy fibers.
In particular, we have straightforwardly shown 
that Fock states of higher order decohere faster 
and that squeezed baths do not help to preserve the quantum
coherence of initially pure number states.
On the other hand, when considering  
coherent superpositions of number states, we have shown 
that squeezed reservoirs can help to slow down 
decoherence, provided that the phase of the bath is optimally 
locked to the coherent phase of the superposition, and that 
the intensity of the squeezing is properly chosen. 
This suggests, looking towards practical implementations, 
that feedback schemes (simulating squeezed reservoirs 
by quantum non demolition measurements) could indeed be 
helpful in prolonging the lifetime of coherent superpositions
of Fock states, i.e. number cat-like states. 
Moreover, it is worth noticing that the same effect on 
the preservation of the purity of the superposition, warranted
by squeezing the bath, can be obtained by an opposite squeezing 
of the initial state. In thermal baths, a superposition 
of squeezed number states 
proves to be more robust against decoherence 
than a superposition of non-squeezed number states.
%%%%%%%%%%%%%%%%%%%%%%%%%%%%%%%%%%%%%%%%%%%%%%%%%
\ack
\vspace*{-.1cm}
We thank INFM, INFN, and MIUR under national project PRIN-COFIN 2002
for financial support.

\end{document}